\documentclass[doublecol]{epl2} 
\usepackage{graphicx}
\usepackage{epstopdf}
\usepackage{bm}% bold math
\usepackage{subfigure}
\usepackage{sidecap}

\usepackage{amsmath}
\newcommand{\bsig}{{\bm\sigma}}

\newcommand{\hbrho}{\hat{{\bm\rho}}}
\newcommand{\hrho}{\hat{\rho}}

\newcommand{\be}{\begin{equation}}
\newcommand{\ee}{\end{equation}}

\newcommand{\bea}{\begin{eqnarray}}
\newcommand{\eea}{\end{eqnarray}}
\newcommand{\bd}{\begin{displaymath}}
\newcommand{\ed}{\end{displaymath}}
\newcommand{\ba}{\begin{array}}
\newcommand{\ea}{\end{array}}
\newcommand{\bi}{\begin{itemize}}
\newcommand{\ei}{\end{itemize}}
\newcommand{\bc}{\begin{center}}
\newcommand{\ec}{\end{center}}
\newcommand{\bfl}{\begin{flushleft}}
\newcommand{\efl}{\end{flushleft}}
\newcommand{\bfr}{\begin{flushright}}
\newcommand{\efr}{\end{flushright}}

\newcommand{\s}{\c{s}}

\newcommand{\CP}{CePt$_3$Si}

\def\ua{\uparrow}
\def\da{\downarrow}

\def\br{{\bf r}}
\def\bk{{\bf k}} \def\bq{{\bf q}}  \def\hg{\rm\hat{g}}
\def\bg{{\bf g}} \def\hbg{\hat{{\bf g}}} \def\bd{{\bf d}}  \def\bS{{\bf S}}
\def\bB{{\bf B}} \def\bS{{\bf S}}
\def\da{\downarrow} \def\ua{\uparrow} 

\def\6{\partial}
  
   \def\e{\epsilon}

\def\s{\sigma}

\def\={\!\!\!&=&\!\!\!}
\def\+{\!\!\!&&\!\!\!+~}
\def\-{\!\!\!&&\!\!\!-~}

\title{Rashba spin-orbit coupling effects in quasiparticle interference of non-centrosymmetric superconductors}
\author{Alireza Akbari%\inst{1} 
\and Peter Thalmeier%\inst{1}
}
\shortauthor{A. Akbari \etal}

\institute{                    
   Max-Planck Institute for the  Chemical Physics of Solids, D-01187 Dresden, Germany \\
  }

\abstract{
The theory of quasiparticle interference (QPI)  for non-centrosymmetric (NCS) superconductors with Rashba spin-orbit coupling 
is developed using T-matrix theory in Born approximation.
We show that qualitatively new effects in the QPI pattern originate from the  Rashba spin-orbit coupling: The resulting
spin coherence factors lead to a distinct difference of charge- and spin- QPI and to an induced spin anisotropy in the
latter even for isotropic magnetic impurity scattering. In particular a cross - QPI appears describing the spin oscillation pattern due to nonmagnetic impurity scattering which is directly related to the Rashba vector.
We apply our theory to a 2D model for the NCS heavy fermion unconventional superconductor CePt$_3$Si and discuss the new QPI features for a gap model with accidental node lines due to its composite singlet-triplet nature. 
}

\begin{document}

\maketitle

The determination of gap symmetry in unconventional superconductors (SC) is a persistent problem. 
Most useful methods for its investigation are ARPES experiments \cite{okazaki:12}, specific heat and thermal transport measurements
under rotating field \cite{matsuda:06}  and  the use of STM-based quasiparticle interference technique (QPI) which utilizes the ripples in electronic density generated by random surface impurities \cite{byers:93,Wang:2003,McElroy:2003,Hanaguri:2009,Maltseva:2009,Hanaguri:2010,Chuang:2010,Akbari:2010,Knolle:2010}. 
In the cuprate
and iron pnictides
 the former is readily applicable but sofar not for heavy fermion unconventional superconductors where SC gaps are only in the meV range.  In this case the last two methods are more powerful. QPI technique has recently been proposed to discriminate between different d-wave pairing states in 115 systems \cite{akbari:11}  and has been successfully demonstrated for CeCoIn$_5$ \cite{allan:13}. Before it was also used to investigate the hidden order state of URu$_2$Si$_2$ \cite{aynajian:10,schmidt:10,yuan:12}.

Here we propose the application of QPI to inversion symmetry-breaking, non-centrosymmetric (NCS) superconductors, notably the tetragonal heavy fermion 131 and 113 compounds \cite{settai:07} like CePt$_3$Si \cite{bauer:07} and CeRhSi$_3$ \cite{kimura:07}. We develop the QPI theory for the case of mixed singlet-triplet gap under the presence of Rashba-type spin-orbit coupling. We show that a wealth of new QPI features is to be expected: i) Distinct differences in the charge- and spin- QPI due to the effect of Rashba coherence factors. ii) Likewise Rashba - induced anisotropies in the spin- QPI even for isotropic magnetic impurity scattering. iii) A new kind of cross- QPI where scattering by {\it nonmagnetic} (charge) impurities leads to spin density pattern directly related to the non-zero Rashba vector (and vice versa). Finally the expected gap nodes in NCS superconductors are generally not on symmetry positions but determined by the ratio of singlet and triplet amplitudes. QPI can give important  information on the position of these accidental nodes. We use a weak coupling BCS theory for the NCS superconductor with a 2D model Fermi surface appropriate for 131 compounds. We employ Nambu Green's function technique and T-matrix theory in Born approximation to calculate the quasiparticle interference spectra and discuss their new features as compared to centrosymmetric unconventional superconductors. 

The present BCS model  is given by \cite{frigeri:06,eremin:06,fujimoto:07}
\bea
\cal{H}_{SC}&=&\sum_{\bk\s\s'}
\bigl[(\varepsilon_\bk-\mu)\s_0+\bg_\bk\cdot\bsig\bigr]_{\s\s'}c_{\bk\s}^\dagger c_{\bk\s'}+\nonumber\\ 
&&\frac{1}{2}\sum_{\bk\s\s'}(\Delta_\bk^{\s\s'}c_{-\bk\s}^\dagger c_{\bk\s'}^\dagger +H.c.).
\label{eq:HBCS}
\eea
Here $\varepsilon_\bk=2t_1(\cos k_x+\cos k_y)+4t_2 \cos k_x \cos k_y$ is the kinetic energy with respect to chemical potential $\mu$ where
$t_1$ are nearest and $t_2$ next nearest neighbor hopping. Furthermore $\bg_\bk=-\bg_{-\bk}$ 
defines the antisymmetric Rashba term due to broken inversion symmetry \cite{samokhin:04}. Therefore the superconducting $2\times 2$ gap
matrix  $\Delta_\bk=[\psi_\bk\s_0+\bd_\bk\cdot\bsig]i\s_y$ has an even singlet ($\psi_\bk$) as well as odd triplet ($\bd_\bk$) components. The latter
must be aligned with the Rashba vector $\bd_\bk=\phi_0\bg_\bk$ to avoid destruction by pairbreaking \cite{sigrist:09}. Here $\bsig=(\sigma_x,\sigma_y,\sigma_z)$ denotes the Pauli matrices. Diagonalization of the model
leads to an effective two band superconductor on the Rashba split bands $(\xi=\pm1)$ given by $\e_{\bk\xi}=\varepsilon_\bk-\mu+\xi |\bg_\bk|$ that have split Fermi surfaces (FS) and opposite helical spin polarizations as well as different superconducting gaps $\Delta_{\bk\xi}=\psi_\bk+\xi |\bd_\bk|$.
For concreteness we employ a 2D model for the tetragonal Ce- based 131 compounds \cite{takimoto:08,takimoto:09}. The possible influence of magnetic order \cite{yanase:08} is not discussed here.
%
%%%%%%%%%%%%%%%%%%%%%% figure %%%%%%%%%%%%%%%%%%%%%%%%%%%%%%%%%%%%%%%%
\begin{figure}%[b]
\centerline{
\hspace{0.12cm}
\includegraphics[width=0.5\linewidth]{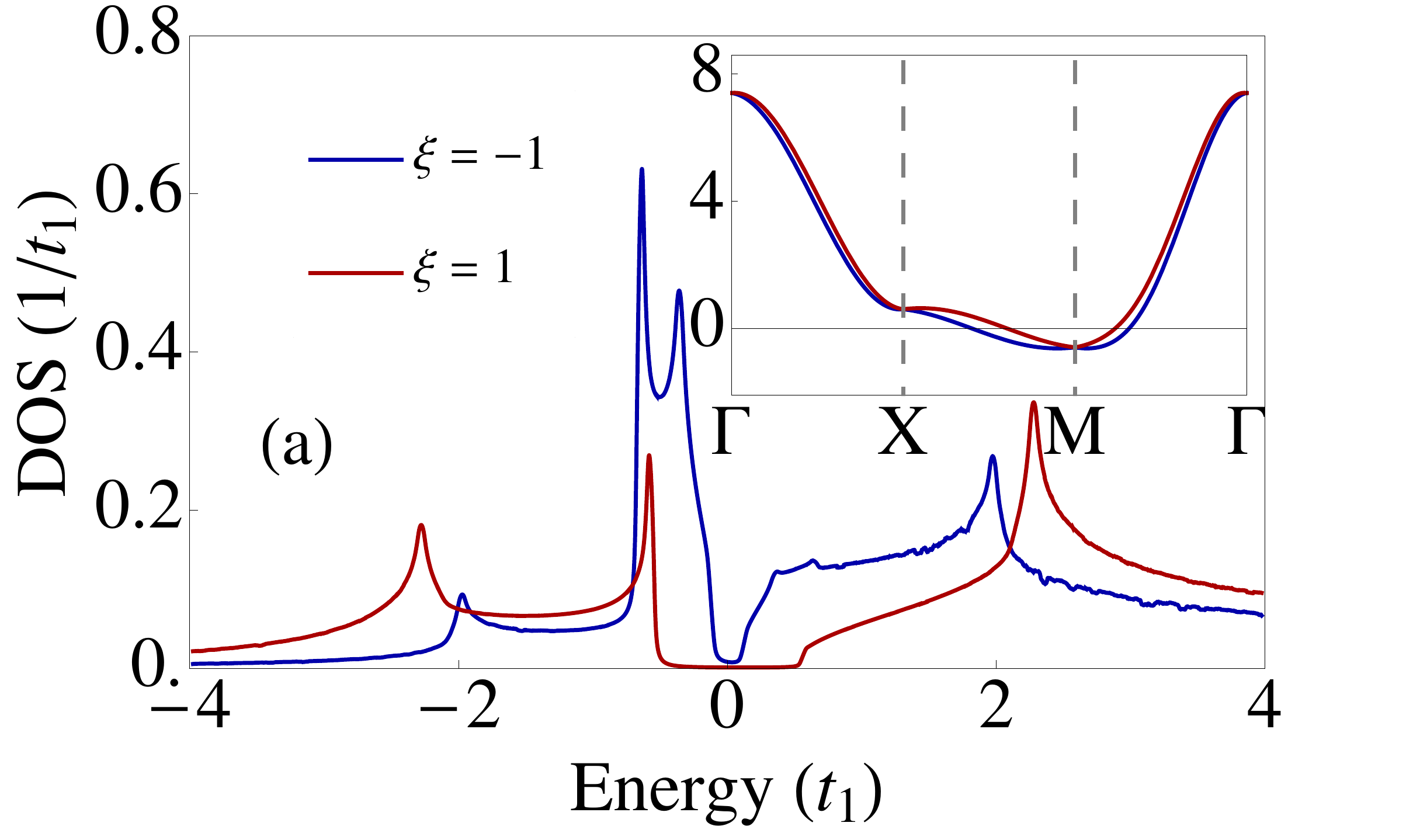}
\hspace{0.14cm}
\includegraphics[width=0.28\linewidth]{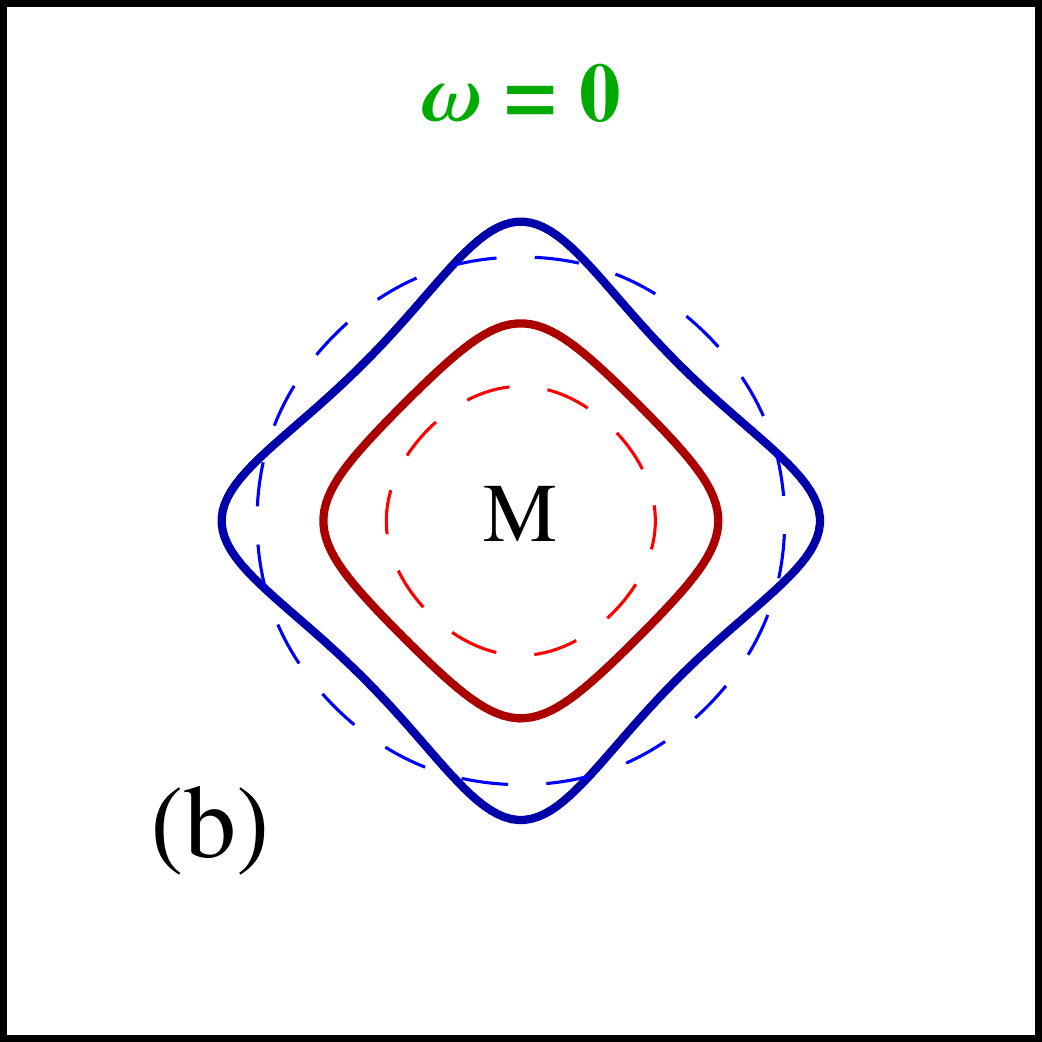}}
\centerline{
\includegraphics[width=0.93\linewidth]{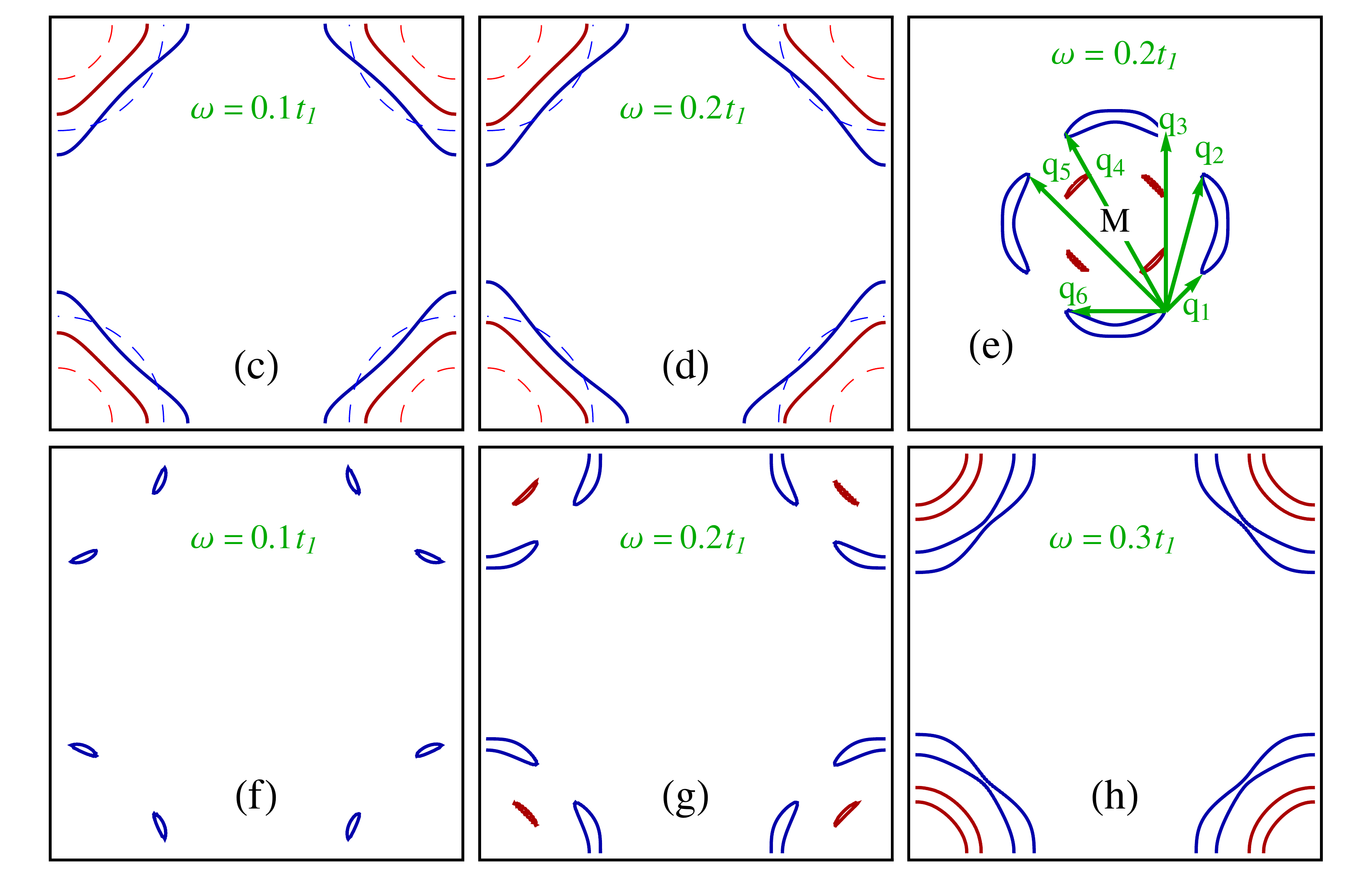}
}
\caption{
a) Electronic density of states (DOS) of Rashba bands ($\xi=\pm1$) in the SC state.  (inset: 2D band structure along $\Gamma$XM$\Gamma$ with $ t_2=0.35t_1, g=0.2t_1$, and $\mu=-2t_1$). Total band width is $W\simeq 8t_1 \equiv T^*\simeq 14$ K (exp. Kondo temperature from \cite{bauer:07}).
b) Normal state electron Fermi surface around M point. Dashed lines indicate nodes of gap functions $\Delta_{\bk\xi}$ with parameters: $\psi_0=2t_1$, $\psi_1=t_1$, and $\phi_0=0.65t_1$ (also in subsequent figures).
c) and d) show spectral functions in the normal state for different energies $\omega$.
e)  Spectral function for the superconducting state around M point.
$\bq_i$ are prominent scattering vectors defining the QPI patterns (c.f. Figs.\ref{Fig3}e, \ref{Fig4}f).
f-h)  Spectral function for the superconducting state  for  different $\omega$.
}
%}
\label{Fig1}
%\vspace{-0.71cm}
%
\end{figure}
%%%%%%%%%%%%%%%%%%%%%%fig%%%%%%%%%%%%%%%%%%%%%%%%%%%%%%%%%%%%%%%%%%5
%

In this class $\bg_\bk=g(\sin k_y,-\sin k_x, 0)$ is in the tetragonal plane. The resulting Rashba-split bands (inset) and density of states (DOS) (in the SC state) are shown in Fig.~\ref{Fig1}a and the electron-type Fermi surface around the M$(\pi,\pi)$ points in Fig.~\ref{Fig1}b. The parameters are chosen \cite{takimoto:08} (caption of Fig.~\ref{Fig1}) to obtain a realistic M-point Fermi surface sheet appropriate for \CP.
The split constant-energy surfaces ($\omega > 0$) are presented in Fig.~\ref{Fig1}b,c. It is known from thermal conductivity \cite{izawa:05} that the superconducting gap has line nodes. To achieve nodes on the M-point Fermi surface we use an extended s-wave \cite{sigrist:09} form for $\psi_\bk$ and $\bd_\bk$ as before. This leads to the two distinct gaps 
\bea
\Delta_{\bk\xi}=\psi_0+\psi_1(\cos k_x+\cos k_y)+\xi \phi_0g_0(\sin k_x^2+\sin k_y^2)^\frac{1}{2}
\eea
on the Rashba-split Fermi surfaces $\e_{\bk\xi}=\mu$ with quasiparticle energies $E_{\bk\xi}=[\epsilon_{\bk\xi}^2+\Delta_{\bk\xi}^2]^\frac{1}{2}$. The gap zeroes are shown as dashed lines in Fig.~\ref{Fig1}b-d. Nodes (node lines in 3D) appear on the FS for the $E_{\bk -}$  quasiparticle sheet but not for $E_{\bk +}$ . Their position is accidental, i.e. determined by the fine tuning of singlet and triplet amplitudes $\psi_1$ and $\phi_0$. 
It was shown in Refs. ~\cite{iniotakis:07,tanaka:09,sato:09} that the nodal case is also topologically nontrivial with possible formation of Andreev bound states (ABS) appearing as surface or edge states. They exist when the surface normal is perpendicular to the Rashba or \bd -vector, i.e., lies within the tetragonal plane and then they lead to a zero-bias conductance peak in Andreev tunneling current.
Here we investigate the opposite case of a tetragonal surface plane with normal parallel to the Rashba vector. Then ABS will not appear and have no signature in the tunneling current \cite{iniotakis:07} or QPI.
We also will not discuss the nodeless gap case with dominating singlet part. It does not correspond to \CP~ and because both Rashba FS sheets are fully gapped have no interesting QPI features in the SC state.
For the nodal gap situation the evolution of constant energy surfaces in the SC state is shown in Fig.~\ref{Fig1}e-h. A few wave vectors $\bq_i$ connecting high curvature points close to nodal positions that will appear prominently in QPI spectra for small $\omega$ are shown in Fig.~\ref{Fig1}e. For larger $\omega$  (Fig.~\ref{Fig1}h) connected FS sheets reappear.
%
%%%%%%%%%%%%%%%%%%%%%% figure %%%%%%%%%%%%%%%%%%%%%%%%%%%%%%%%%%%%%%%%%%%%
\begin{figure}[t]
\centerline{
\includegraphics[width=0.98\linewidth]{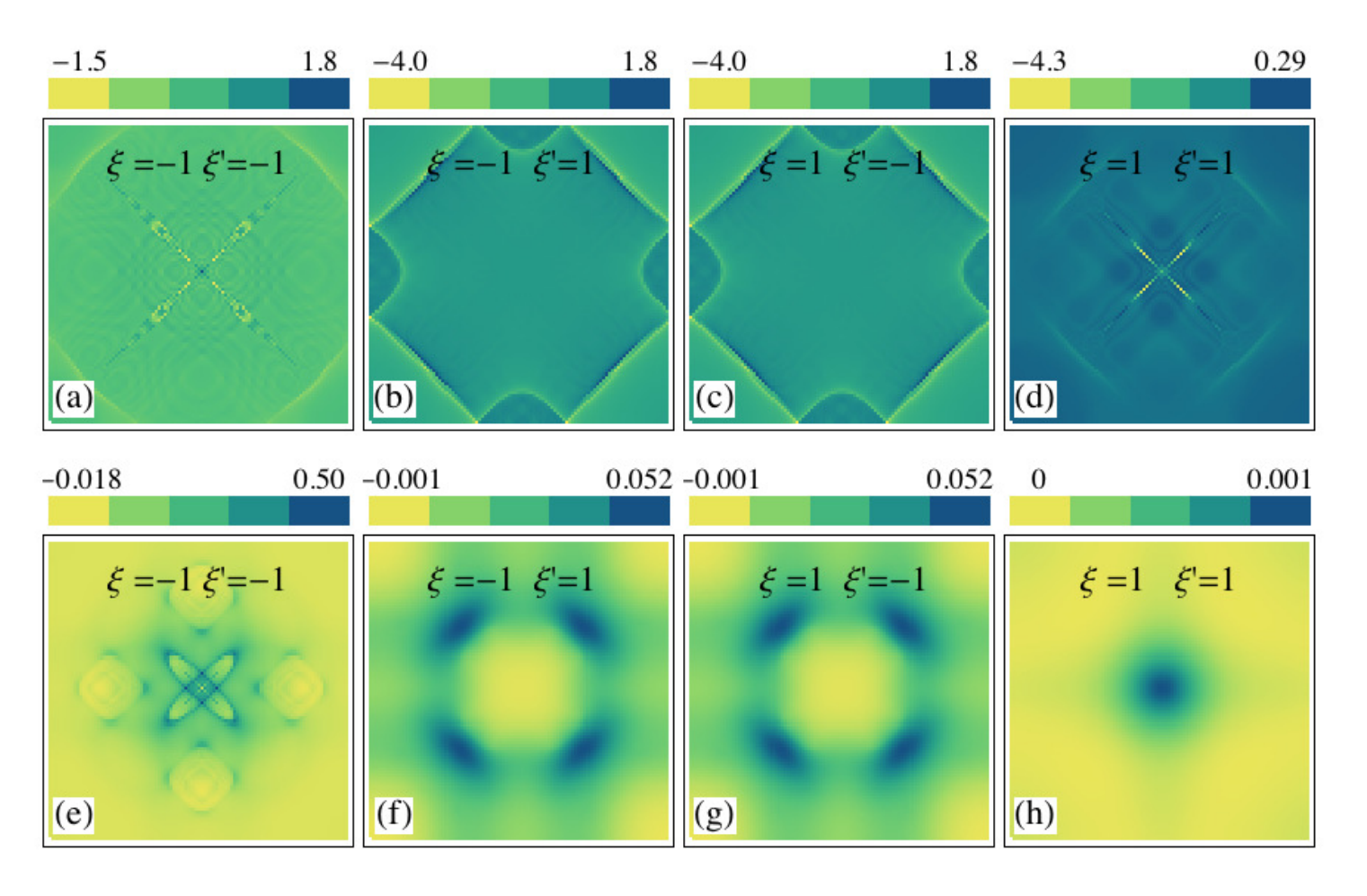}
}
\caption{
The individual  charge- QPI ($\Lambda^\bq_{00}$) contributions for intra- or inter- band scattering with $\xi,\; \xi^\prime=\pm 1$,  from normal (non-magnetic) impurities at the energy $\omega=0.2t_1$:
a-d)  in the normal state 
e-h)  in the superconducting state.
(Summation over each of the rows is shown in Fig.\ref{Fig3}.b and Fig.\ref{Fig3}.e).
}
%}
\label{Fig2}
\end{figure}
%%%%%%%%%%%%%%%%%%%%%%fig%%%%%%%%%%%%%%%%%%%%%%%%%%%%%%%%%%%%%%%%%%%%%%%
%
For the calculation of the QPI pattern we use the normal and anomalous $2\times 2$ spin-space matrix Greens functions of the unperturbed system
$G=G_+\s_0+G_-(\hbg_\bk\cdot\bsig)$ and $F=[F_+\s_0+F_-(\hbg_\bk\cdot\bsig)]i\sigma_y$, respectively, where the scalar Green's functions are given by ($\hbg_\bk=\bg_\bk/|\bg_\bk|$) 
\bea
     \begin{aligned}
G_\pm(\bk,i\omega_n)&=\frac{1}{2}\sum_\xi 
\left\{ \begin{array}{rrrr}
1\\
\xi \end{array} \right\}
\frac{(i\omega_n+\e_{\bk\xi})}{[(i\omega_n)^2-E^2_{\bk\xi}]},
%\nonumber
\\
F_\pm(\bk,i\omega_n)&=\frac{1}{2}\sum_\xi 
\left\{ \begin{array}{rrrr}
1\\
\xi \end{array} \right\}
\frac{\Delta_{\bk\xi}}{[(i\omega_n)^2-E^2_{\bk\xi}]}.
\label{eq:Green}
 \end{aligned}
\eea
The QPI density oscillations are obtained from the full Green's function
that  is determined by the effect of  scattering from random charge and spin impurities at the surface. The total scattering Hamiltonian  in compact form reads
\bea
{\it H}_{imp}=\sum_{\bk\bq\alpha}V_\alpha(\bq)S_\alpha\Psi_{\bk+\bq}^\dagger\hat{\rho}_\alpha\Psi_\bk.
\label{eq:impurity}
\eea
Here we use the Nambu 4-component spinor representation $\Psi^\dagger=(c^\dagger_{\bk\ua} c^\dagger_{\bk\da} c_{-\bk\ua} c_{-\bk\da})$. In the Nambu space the $4\times4$ matrices $\hrho_\alpha$ ($\alpha =(0,i) =(0,x,y,z)$) are given by  $\{\hrho_\alpha\}=(\hrho_0,\hbrho) = (\tau_3\sigma_0,\tau_0\sigma_x,\tau_3\sigma_y,\tau_0\sigma_z)$. The $\tau_\alpha$- and $\sigma_\alpha$- Pauli matrices (with $\tau_0=\sigma_0\equiv 1$) act on Nambu and spin indices, respectively. Furthermore we define $\{S_\alpha\}=(1,\bS)$. The first index $\alpha=0$ corresponds to nonmagnetic impurity scattering $V_0(\bq)$ and entries $i=x,y,z$ to  isotropic magnetic exchange scattering $V_i(\bq)=V_{ex}(\bq)$  from impurity spins \bS. Their components $S_i$ are treated as frozen, i.e. polarized in a given direction by a small magnetic field. An important consequence of the Rashba term is that spin and charge channels for impurity scattering are not decoupled, this also holds true for spin and charge response functions \cite{takimoto:08}.

\begin{figure}%[b]
\centerline{
\includegraphics[width=0.96\linewidth]{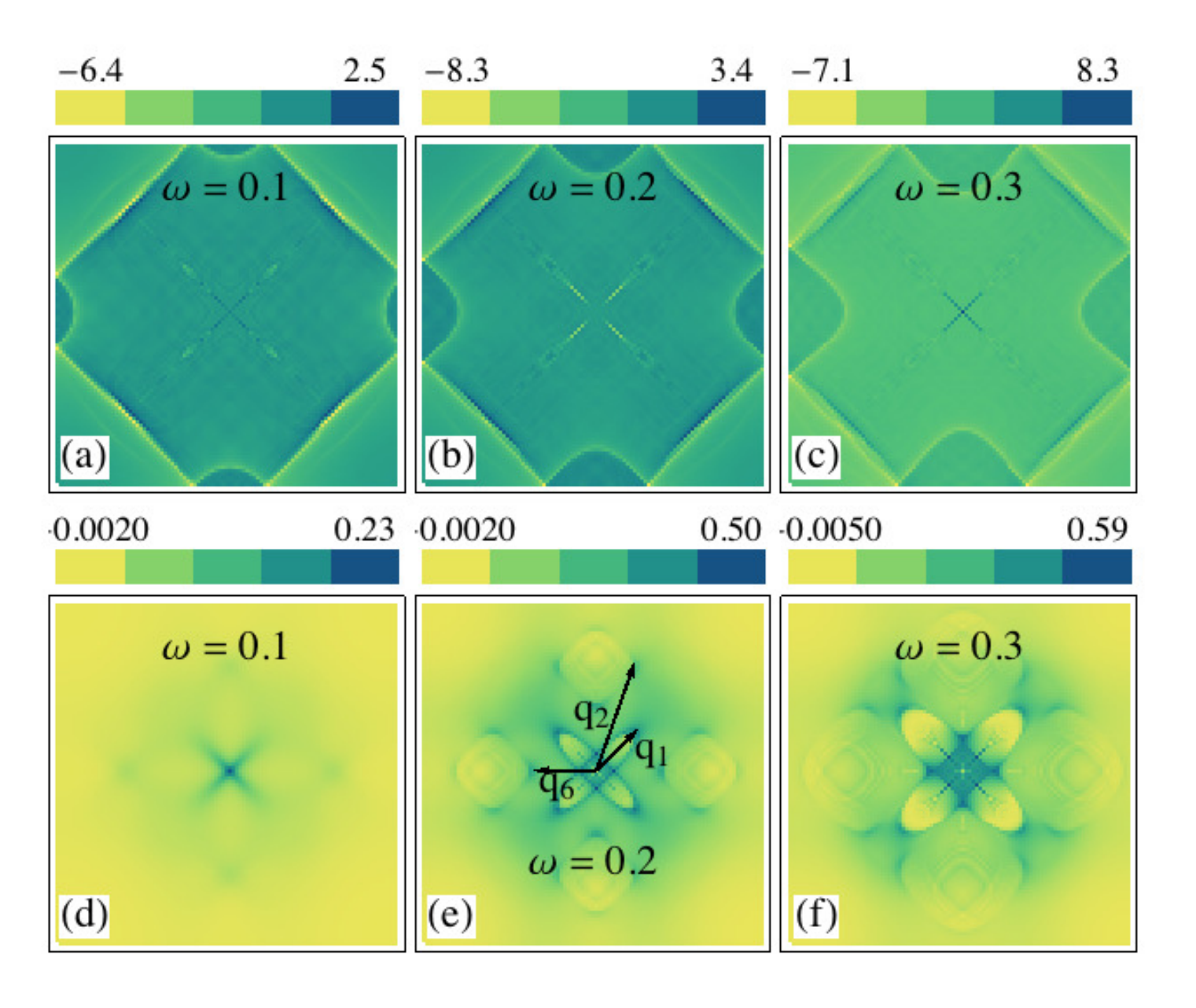}
}
\caption{
a-c) Total charge- QPI ($\Lambda^\bq_{00}$) in the normal state for different $\omega$ and  scattering from non-magnetic impurities.
d-f) The same quantity for the superconducting state.
 Subset $(\bq_1,\bq_2,\bq_6)$ of characteristic QPI wave vectors correspond to those in  the spectral functions given in   Fig.\ref{Fig1}.e ($\omega=0.2t_1$). The frequency satisfies $\omega < |\Delta_{\bk\xi}|\simeq 0.5t_1 \ll W = T^*=8t_1$.
}\vspace{-0.31cm}
%}
\label{Fig3}
\end{figure}

We calculate the change in STM tunneling conductance in charge or spin channel $\alpha (0,x,y,z) $ due to impurity scattering in charge or spin channel $\beta (0,x,y,z)$. For magnetic impurity the scattering channel $\beta$ is fixed by applying a small field $H \ll H_{c2}$ along the x,y,z axis. The conductance channel $\alpha$ is selected by using either a nonmagnetic ($\alpha=0$) or a half-metallic (fully spin polarized) tunneling tip with moment polarized along $\alpha =x,y,z$ and an exchange splitting larger than the heavy fermion quasiparticle band width. Such configuration in principle would allow to determine all elements of the QPI differential conductance tensor. It is given by \cite{byers:93}
\bea 
     \begin{aligned}
\frac{d\delta I_\alpha(\br,V)}{dV}
&\sim
 \delta N_{\alpha\beta}(\br,\omega = V)
\\
&=-\frac{1}{\pi}{\rm Im}[{\rm Tr}_\sigma\hrho_\alpha\delta \hat{G}_\beta(\br,\br,\omega)]_{11}.
%\nonumber
     \end{aligned}
\eea
Here $\delta \hat{G}_\beta$ is the change of the $4 \times 4$ matrix Green's function in combined Nambu and spin space (each with dimension 2) which is due to impurity scattering in charge or spin channel $\beta (0,x,y,z)$. Furthermore matrix index (11) refers to the Nambu space which results from the trace with respect to $\tau$ including the projector $\frac{1}{2}(1+\tau_z)$. The remaining trace refers to spin space only.
In this work we focus on  the spatial oscillations or momentum dependence by weak scattering and ignore the possibility of bound state formation \cite{liu:08} in the strong scattering limit. Therefore we treat the former in Born approximation which leads to the Fourier transform of differential conductances given by  $(\bk'=\bk-\bq)$
\be
     \begin{aligned}
\delta N_{\alpha\beta}(\bq,\omega)=&-\frac{1}{\pi}V_\beta(\bq)
 {\rm Im} 
 {\Big [}
 \Lambda_{\alpha\beta}(\bq,i\omega_n)
 {\Big]}_{i\omega_n\rightarrow \omega + i\delta},
\\
\Lambda_{\alpha\beta}(\bq,i\omega_n)=&
\frac{1}{N}\sum_{\bk}
{\rm Tr}_{\s}
{\Big [}
\hrho_\alpha\hat{G}(\bk ,i\omega_n)\hrho_\beta\hat{G}(\bk',i\omega_n)
{\Big ]}_{11},
%\nonumber
\label{eq:Born}
     \end{aligned}
\ee
where $N=L^2$ is the number of grid points. We assume here that the tunneling happens out of the coherent
heavy quasiparticle states. This is justified for temperature $T$  and frequency $\omega$ much smaller than the Kondo temperature $T^*$ \cite{schmidt:10} which is of the order 14 K for \CP~\cite{bauer:07}. We therefore restrict to frequencies (Fig.~\ref{Fig3}) of the order of the SC gap and we do not intend to describe the Fano resonance shape that appears for higher frequencies of the order of the effective quasiparticle bandwidth $T^*$ \cite{schmidt:10}.
The QPI function $\Lambda_{\alpha\beta}(\bq,i\omega_n)$ is evaluated using Eq.~(\ref{eq:Green}),  performing the $\sigma-$ trace we obtain per spin
\begin{equation}
     \begin{aligned}
\label{eq:Lambda}
\Lambda^\bq_{00}(i\omega_n)=&\frac{1}{4N}\sum_{\bk\xi\xi'}
\bigl[1+\xi\xi'(\hbg_\bk\cdot\hbg_{\bk'})\bigr]K^{\bk\bq}_{\xi\xi'}(i\omega_n),
\\
\Lambda^\bq_{\rm ii}(i\omega_n)=&
\frac{1}{4N}
\sum_{\bk\xi\xi'}
\bigl[1-\xi\xi'(\hbg_\bk\cdot\hbg_{\bk'}-2{\hg}^{\rm i}_\bk{\hg}^{\rm i}_{\bk'})\bigr]K^{\bk\bq}_{\xi\xi'}(i\omega_n),
\\
\Lambda^\bq_{{\rm i}0}(i\omega_n)=&\frac{1}{2N}\sum_{\bk\xi\xi'}
\xi{\hg}^{\rm i}_\bk K^{\bk\bq}_{\xi\xi'}(i\omega_n),
     \end{aligned}
\end{equation}
where the integration kernel for intra-band ($\xi=\xi'$) and inter-band ($\xi\neq\xi'$) processes is given by
\be
K^{\bk\bq}_{\xi\xi'}(i\omega_n)=\frac
{(i\omega_n+\e_{\bk\xi})(i\omega_n+\e_{\bk-\bq\xi'})-\Delta_{\bk\xi}\Delta_{\bk-\bq\xi'}}
{[(i\omega_n)^2-E^2_{\bk\xi}] [(i\omega_n)^2-E^2_{\bk-\bq\xi'}]}.\nonumber
\label{eq:kernel}
\ee
%
%%%%%%%%%%%%%%%%%%%%%% figure %%%%%%%%%%%%%%%%%%%%%%%%%%%%%%%%%%%%%%%%%%%%%
%%%%%%%%%%%%%%%%%%%%%%fig%%%%%%%%%%%%%%%%%%%%%%%%%%%%%%%%%%%%%%%%%%%%%%%%
%
These equations give the  QPI patterns for the NCS superconductors.
 Before presenting numerical results we point out the salient new features
as compared to centrosymmetric unconventional superconductors \cite{capriotti:03}.
Here $\Lambda_{00}$ describes the charge- QPI pattern due
to nonmagnetic impurities, $\Lambda_{ii}$ the (diagonal) spin- QPI polarization pattern due to magnetic impurities and
$\Lambda_{i0}$ the spin polarization generated by {\it nonmagnetic} impurities which we term as cross- QPI. It occurs only for finite Rashba coupling $\bg_\bk$.
The coupling of charge and spin degrees in cross- QPI appears because the Rashba band states are described by helicity and not spin quantum numbers. Therefore in general a particle-hole (charge) excitation will also change the spin state. The diagonal
QPI functions in Eq.~(\ref{eq:Lambda}) are finite even for $\bg_\bk=0$. However they are strongly modified when inversion symmetry is broken due to the Rashba helicity coherence factors in square brackets of Eq.~(\ref{eq:Lambda}) . The latter have different sign for pure charge and spin QPI functions and also change sign between intra- and interband contributions. The coherence factors can amplify or annihilate the individual contributions in the sum depending on the relative orientation of Rashba vectors and band indices.
Therefore one expects that charge- ($\Lambda_{00}(\bq)$) and spin- ($\Lambda_{ii}(\bq)$) QPI pattern are profoundly different as opposed to non-Rashba case where they should be identical. The spin- QPI pattern shows a further striking Rashba effect: Although the exchange scattering itself is isotropic,  $\Lambda_{ii}(\bq)$ is to be expected anisotropic in the present case when $\bg_\bk$ is confined to the tetragonal plane as seen from  Eq.~(\ref{eq:Lambda}). Furthermore the Rashba term leads to nondiagonal elements in the spin- QPI pattern, in the present case to $\Lambda^\bq_{xy}(i\omega_n)=\Lambda^\bq_{yx}(i\omega_n)$ which is given by
\be
\Lambda^\bq_{xy}(i\omega_n)=\frac{1}{4N}\sum_{\bk\xi\xi'}
\xi\xi'(\hg^x_\bk\hg^y_{\bk-\bq}+\hg^y_\bk\hg^x_{\bk-\bq}) K^{\bk\bq}_{\xi\xi'}(i\omega_n).
\label{eq:Lambdaxy}
\ee
The QPI functions $\Lambda^\bq_{00}$ and  $\Lambda^\bq_{ij}$ are even and $\Lambda^\bq_{i0}$ is odd in \bq. Therefore $\delta N_{i0}(\br,\omega)$ is finite only when $V_0(\bq)$ contains odd (p-wave) contributions and the net (area integrated) cross-QPI conductance always vanishes.

We conclude that QPI for non-centrosymmetric superconductors derived here exhibits a wealth of new effects due to the inversion symmetry breaking Rashba term which we present now for the 2D model of 131 compounds.\\
%
%%%%%%%%%%%%%%%%%%%%%% figure %%%%%%%%%%%%%%%%%%%%%%%%%%%%%%%%%%%%%%%%%%%
\begin{figure}%[h]
\centerline{
\includegraphics[width=0.96\linewidth]{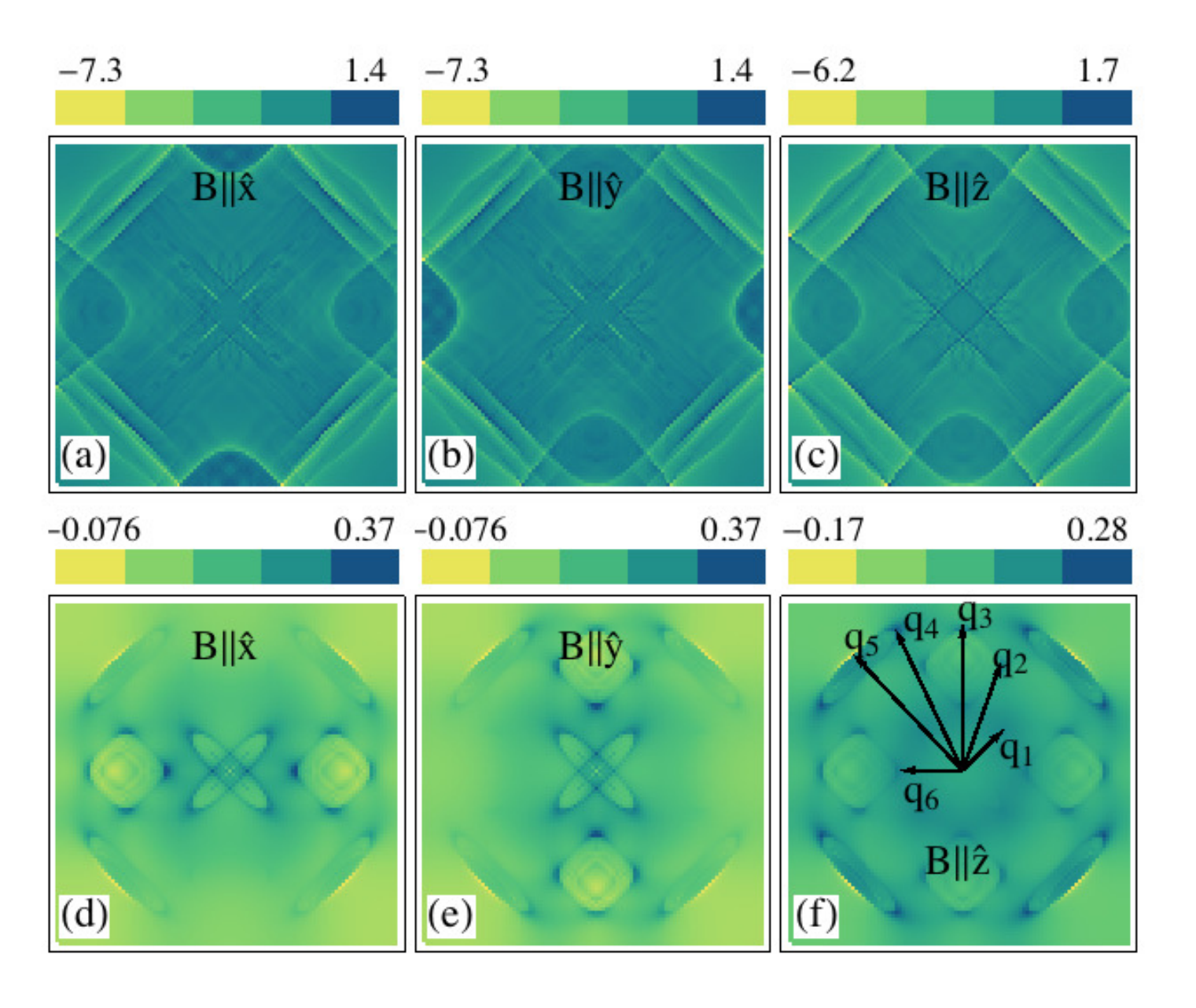}
}
\caption{
a-c) Total spin- QPI ($\Lambda^\bq_{ii}$) for the normal state  for exchange scattering from magnetic impurities at energy $\omega=0.2t_1$.
d-f) The same quantity for the superconducting state.
Each panel corresponds to different orientation of the impurity spin \bS~ parallel to an applied field $\bB \rightarrow 0$.
Characteristic QPI wave vectors $\bq_i$ (Fig.\ref{Fig1}.e) are shown for $\bB \parallel z$.
} 
\label{Fig4}
\end{figure}
%%%%%%%%%%%%%%%%%%%%%%fig%%%%%%%%%%%%%%%%%%%%%%%%%%%%%%%%%%%%%%%%%%%%%%
%%%%%%%%%%%%%%%%%%%%%%% figure %%%%%%%%%%%%%%%%%%%%%%%%%%%%%%%%%%%%%%%%%%%
\begin{figure}
\centering
\includegraphics[width=0.65\linewidth]{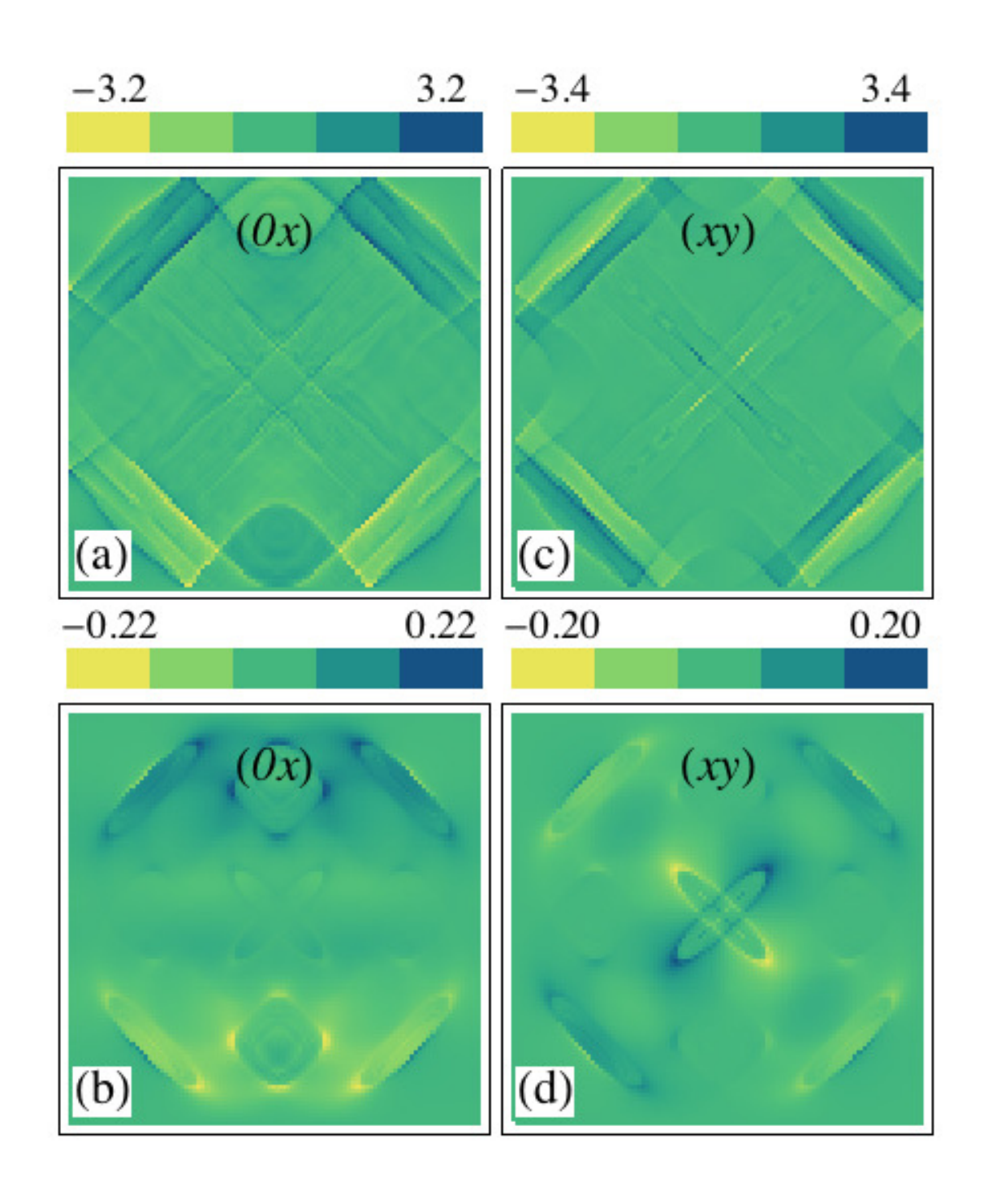}
\caption{
a) Total cross- QPI of odd spin density $\Lambda^\bq_{0x}$ in the normal state from {\it non-magnetic} scattering, and
b) same quantity in superconducting state.
c) Total non-diagonal even spin- QPI $\Lambda^\bq_{xy}$ for the normal state  from exchange scattering 
by magnetic impurities, and
d) same quantity in superconducting state. Here $\bB\rightarrow 0$ is $\parallel$ y and $\omega=0.2t_1$. 
}
%}
\label{Fig5}
\end{figure}
%%%%%%%%%%%%%%%%%%%%%%fig%%%%%%%%%%%%%%%%%%%%%%%%%%%%%%%%%%%%%%%%%%%%%%
%
%
All QPI spectra have four contributions from two intraband and two (equivalent) interband scattering processes. They are shown individually in Fig.\ref{Fig2}a-d for charge-QPI in the normal state and in Fig.\ref{Fig2}e-h for the SC state. In the former the scattering {\it across} the Fermi surface of Fig.~\ref{Fig1}b maps out the '2k$_F$' contours, partly folded back into the first BZ. They are slightly different due to the different diameter of Rashba split sheets. The diagonal center cross  appears only for intraband contributions due to scattering process {\it parallel} to each Rashba sheet.
In the SC state for small $\omega$ the sheets break up into small pieces in BZ regions connecting the node points where the gap is small. Therefore the $E_{\bk -}$ Rashba band dominates because it is the only one with nodes  (Fig.\ref{Fig1}b,e) and the SC QPI spectrum reflects the $(\xi,\xi')=(-1,-1)$ intraband transitions in Fig.\ref{Fig2}e. A set of the typical wave vectors $\bq_i$ of Fig.\ref{Fig1}e selected by the coherence factors show up as prominent spots in Fig.\ref{Fig2}e. The other three contributions have lower amplitude due to the small $E_{\bk +}$ sheets at this energy. The sum of all four contributions is the  total charge-QPI spectrum which is presented in Fig.\ref{Fig3}a-f for the normal and SC state for three energies (bias voltages V): $\omega /t_1= 0.1,\; 0.2$, and $0.3$. In the former (a-c) the '$2k_F$' contours which increase with $\omega$ are now blurred due contributions with slightly different dimensions. The diagonal cross due to inraband processes appears at all energies. In the SC state for small $\omega$ the set of prominent spots at $\bq_i$  due to  Fig.\ref{Fig2}e survive and are shown explicitly in Fig.\ref{Fig3}e. The observation of these features in QPI and their variation with  $\omega$ would allow to confirm directly the gap model with node structure in Fig.\ref{Fig1}b,e. For larger $\omega$ instead of spots at $\bq_i$ arcs appear due to scattering on and between the again connected constant-$\omega$ surfaces of both Rashba bands in Fig.\ref{Fig1}h.

In the centrosymmetric superconductor $(g_\bk =0)$ within Born approximation the charge- and spin- QPI functions in Eq.~(\ref{eq:Lambda}) are identical. The presence of a Rashba term introduces two new aspects due to the appearance of coherence factors (square brackets in Eq.~(\ref{eq:Lambda})): i) the charge-and spin-QPI becomes different due to different intra-band contributions in the two cases ii) While the charge- QPI retains the fourfold symmetry as obvious from the first of Eq.~(\ref{eq:Lambda}) and Fig.~\ref{Fig3}, the spin-QPI  coherence factors in the second Eq.~(\ref{eq:Lambda}) explicitly breaks rotational symmetry with respect to the impurity moment direction.
This is shown in  Fig.\ref{Fig4} for normal and SC phase where we assume that a small magnetic field \bB~ polarizes the impurity spins \bS~ along one of the three symmetry directions. The spin- QPI pattern for \bB~ along x,y directions are still equivalent but rotated by $90^\circ$, however, both are quite different from the case $\bB \parallel z$, in particular in the SC state. This is because in the present model the Rashba vector fulfils $\bg_\bk\cdot\hat{{\bf z}}=0$ which singles out the z direction. This anisotropy of spin- QPI is therefore characteristic for non-centrosymmetric superconductors. The spin-QPI pattern is dominated by the full set of  scattering vectors $\bq_i$ connecting the high-curvature points of the constant-$\omega$ surface (Fig.\ref{Fig4}f). 

Sofar we have only considered diagonal charge- or spin- QPI, however it is a very peculiar feature of NCS Rashba systems that  charge-spin cross- QPI ( and the reverse) described by $\Lambda^\bq_{i0}$ as well as non-diagonal spin-QPI, e.g. $\Lambda^\bq_{xy}$ exist. The former describes spin oscillations introduced by non-magnetic impurities (and vice versa) which are odd under inversion because $\Lambda^{-\bq}_{i0}=-\Lambda^{\bq}_{i0}$. It is shown in Fig.\ref{Fig5}a,b for (x0) component with spin polarization along x, again the (y0) case is just rotated in the BZ by  $90^\circ$ while it vanishes for (z0). 
Finally we show in Fig.\ref{Fig5}c,d the nondiagonal (xy) spin pattern polarized perpendicular to the impurity spins and described by 
 Eq.~(\ref{eq:Lambdaxy}). Only the (xy) component is non-zero because $\bg_\bk$ lies in the tetragonal plane. It also exhibits
 reflection symmetry with respect to diagonals (interchange of x,y directions).
 It would also be interesting to investigate QPI for the tetragonal NCS in a geometry where the surface normal is within the tetragonal plane. Then Andreev bound states may appear \cite{iniotakis:07} and lead to additonal contributions to QPI spectrum for small bias voltages, similar as in the zero bias conductance peak. To treat this case one has to include the quasiparticle dispersion perpendicular to the tetragonal plane and extend the t-matrix formalism to include inhomogeneous states.
 \\
We have shown that QPI in noncentrosymmetric systems exhibits a wealth of new features due to the presence of Rashba spin orbit coupling. They are important signatures of the helical spin texture and a powerful tool to investigate the nodal structure of the superconducting gap, in particular because the nodal positions are not fixed by symmetry for mixed singlet-triplet gaps.

We thank T. Takimoto for helpful comments and I. Eremin for useful discussion.

%%%%%%%%%%%%%%%%%%%%%%%%%%%%%      References        %%%%%%%%%%%%%%%%%\section*{References}
\bibliography{References}

\bibliographystyle{EPL}

\end{document}